\documentclass[a4paper]{article}
\usepackage{a4wide}
\usepackage{graphicx}
\usepackage{amsmath}
\usepackage{amsthm}
\usepackage{amssymb, latexsym, mathrsfs}
\usepackage{enumerate}
\usepackage{algorithm}
\usepackage{color}
\usepackage{transparent}
\usepackage[affil-it]{authblk}
\usepackage[colorlinks=true,citecolor=blue]{hyperref}
\usepackage{listings}
\newcommand{\NN}{{\mathcal N}}

\newcommand{\subss}[2]{#1_{[#2]}}

\newcommand{\diag}{\mbox{diag}}

\newcommand{\Xset}{\mathbb{X}}

\newcommand{\Uset}{\mathbb{U}}

\newcommand{\abs}[1]{{|{#1}|}}
\newcommand{\norme}[2]{{||{#1}||_{#2}}}

\newcommand{\Pset}{\mathbb{P}}

\newcommand{\ba}[1]{\begin{array}{#1}}
\newcommand{\ea}{\end{array}}

\newcommand{\matr}[1]{
\begin{bmatrix}
    #1
\end{bmatrix}
}

\begin{document}

     \title{Hycon2 Benchmark: Power Network System \thanks{The research leading to these results has received funding from the European Union Seventh Framework Programme [FP7/2007-2013]  under grant agreement n$^\circ$ 257462 HYCON2 Network of excellence.}}
     \author{Stefano Riverso%
       \thanks{Electronic address: \texttt{stefano.riverso@unipv.it};}} 
     \affil{}

     \author{Giancarlo Ferrari-Trecate%
       \thanks{Electronic address: \texttt{giancarlo.ferrari@unipv.it}}} 
     \affil{Dipartimento di Ingegneria Industriale e dell'Informazione\\Universit\`a degli Studi di Pavia\\via Ferrata, 1\\27100 Pavia\\Italy}

     \date{June, 2012}

     \maketitle

     \begin{abstract}
       As a benchmark exercise for testing software and methods developed in Hycon2 for decentralized and distributed control, we address the problem of designing the Automatic Generation Control (AGC) layer in power network systems. In particular, we present three different scenarios and discuss performance levels that can be reached using Centralized Model Predictive Control (MPC). These results can be used as a milestone for comparing the performance of alternative control schemes. Matlab software for simulating the scenarios is also provided in an accompanying file.
     \end{abstract}

     \newpage

     \section{Introduction}
          \label{sec:PNS}
          An example of a real application that can benefit of decentralized and distributed control schemes is the regulation of a Power Network System (PNS). We consider a PNS as composed by several power generation areas coupled through tie-lines \cite{Saadat2002}. The aim is to design the Automatic Generation Control (AGC) layer for frequency control with the goal of: 
          \begin{itemize}
          \item keeping the frequency approximately at the nominal value;
          \item controlling the tie-line powers in order to reduce power exchanges between areas. In the asymptotic regime each area should compensate for local load steps and produce the required power.
          \end{itemize}
          
          We consider thermal power stations with single-stage turbines. The dynamics of an area equipped with primary control and linearized around equilibrium value for all variables can be described by the following continuous-time LTI model \cite{Saadat2002}
          \begin{equation}
            \label{eq:ltipower}
            \subss{\Sigma}{i}^C:\quad\subss{\dot{x}}{i} = A_{ii}\subss x i + B_{i}\subss u i + L_{i}\Delta P_{L_i} + \sum_{j\in\NN_i}A_{ij}\subss x j
          \end{equation}
          where $\subss x i=(\Delta\theta_i,~\Delta\omega_i,~\Delta P_{m_i},~\Delta P_{v_i})$ is the state, $\subss u i = \Delta P_{ref_i}$ is the control input of each area, $\Delta P_{L}$ is the local power load and $\NN_i$ is the sets of neighboring areas, i.e. areas directly connected to $\subss\Sigma i^C$ through tie-lines. The matrices of system \eqref{eq:ltipower} are defined as
          \begin{equation}
            \label{eq:matrixpower}
            \begin{aligned}
              A_{ii}(\{P_{ij}\}_{j\in\NN_i}) &= \matr{ 0 & 1 & 0 & 0 \\ -\frac{\sum_{j\in\NN_i}{P_{ij}} }{2H_i} & -\frac{D_i}{2H_i} & \frac{1}{2H_i} & 0 \\ 0 & 0 & -\frac{1}{T_{t_i}}  & \frac{1}{T_{t_i}} \\ 0 & -\frac{1}{R_iT_{g_i}} & 0 & -\frac{1}{T_{g_i}} }
              &B_{i} = \matr{ 0 \\ 0 \\ 0 \\ \frac{1}{T_{g_i}} }\\
              A_{ij} &= \matr{ 0 & 0 & 0 & 0 \\ \frac{P_{ij}}{2H_i} & 0 & 0 & 0 \\ 0 & 0 & 0  & 0 \\ 0 & 0 & 0 & 0 }
              &L_{i} = \matr{ 0 \\ -\frac{1}{2H_i} \\ 0 \\ 0 }
            \end{aligned}
          \end{equation}
          For the meaning of constants as well as some typical parameter values we defer the reader to Table \ref{tab:networkparameter}.
          \begin{table}[!ht]
            \footnotesize
            \centering
            \begin{tabular}{|c|c|}
              \hline
              $\Delta\theta_i$ & Deviation of the angular displacement of the rotor with respect to the stationary reference axis on the stator \\
              $\Delta\omega_i$ & Speed deviation of rotating mass from nominal value\\
              $\Delta P_{m_i}$ & Deviation of the mechanical power from nominal value (p.u.)\\
              $\Delta P_{v_i}$ & Deviation of the steam valve position from nominal value (p.u.)\\
              $\Delta P_{ref_i}$ & Deviation of the reference set power from nominal value (p.u.)\\
              $\Delta P_{L_i}$ & Deviation of the nonfrequency-sensitive load change from nominal value (p.u.)\\
              $H_i$ & Inertia constant defined as $H_i=\frac{\mbox{kinetic energy at rated speed}}{\mbox{machine rating}}$ (typically values in range $[1-10]\mbox{ sec}$) \\
              $R_i$ & Speed regulation \\
              $D_i$ & Defined as $\frac{\mbox{percent change in load}}{\mbox{change in frequency}}$ \\
              $T_{t_i}$ & Prime mover time constant (typically values in range $[0.2-2]\mbox{ sec }$)\\
              $T_{g_i}$ & Governor time constant (typically values in range $[0.1-0.6]\mbox{ sec }$) \\
              $P_{ij}$ & Slope of the power angle curve at the initial operating angle between area $i$ and area $j$ \\
              \hline
            \end{tabular}
            \caption{Variables of a generation area with typical value ranges \cite{Saadat2002}. (p.u.) stands for ``per unit''.}
            \label{tab:networkparameter}
          \end{table}
          \normalsize

          We note that model~\eqref{eq:ltipower} is input decoupled since both $\Delta P_{ref_i}$ and $\Delta P_{L_i}$ act only on subsystem $\subss{\Sigma}{i}^C$. Moreover, subsystems $\subss\Sigma i^C$ are parameter dependent since the local dynamics depends on the quantities $-\frac{\sum_{j\in\NN_i}{P_{{ij}}} }{2H_i}$. 

          In the following we introduce three scenarios corresponding to different interconnection topologies of generation areas. The model parameters and constraints on $\Delta\theta_i$ and on $\Delta P_{ref_i}$ for systems in all Scenarios are given in Table \ref{tab:scenario123}. We highlight that all parameter values are within the range of those used in Chapter 12 of \cite{Saadat2002}. We define $M$ as the number of areas in the power network. For each scenario, discrete-time models $\subss\Sigma i$ with $T_s = 1$ sec sampling time are obtained from $\subss\Sigma i^C$ using two alternative discretization schemes.
          \begin{itemize}
          \item Exact discretization of the overall system (acronym $D$);
          \item Discretization system-by-system, i.e. exact discretization for each area treating $\subss u i$, $\Delta P_{L_i}$ and $\subss x j,~j\in\NN_i$ as exogenous inputs (acronym $Dss$).
          \end{itemize}
          In particular, we note that $Dss$ preserves the input-decoupled structure of $\subss\Sigma i^C$ while $D$ does not.

          \begin{table}[!ht]
            \centering
            \begin{tabular}{|c|c|c|c|c|c|}
              \hline
              &  Area 1 & Area 2 & Area 3 & Area 4 & Area 5 \\
              \hline
              $H_i$     &  12       & 10        &  8        &  8        & 10         \\
              \hline
              $R_i$     &   0.05   & 0.0625 & 0.08    &  0.08   &  0.05    \\
              \hline
              $D_i$     &  0.7      & 0.9      & 0.9      &  0.7     &  0.86     \\
              \hline
              $T_{t_i}$ &   0.65    & 0.4      & 0.3      &  0.6     &  0.8       \\
              \hline
              $T_{g_i}$ &  0.1      & 0.1      & 0.1      &  0.1     &   0.15    \\
              \hline
              \multicolumn{6}{c}{}                                                          \\
            \end{tabular}
            
            \begin{tabular}{|c|c|c|c|c|c|}
              \hline
              &  Area 1 & Area 2 & Area 3 & Area 4 & Area 5 \\
              \hline
              $\Delta\theta_i$    &  $\norme{\subss{x}{1,1}}{\infty}\leq 0.1$   &  $\norme{\subss{x}{2,1}}{\infty}\leq 0.1$   &  $\norme{\subss{x}{3,1}}{\infty}\leq 0.1$ &   $\norme{\subss{x} {4,1}}{\infty}\leq 0.1$   &  $\norme{\subss{x}{5,1}}{\infty}\leq 0.1$    \\
              \hline
              $\Delta P_{ref_i}$    &  $\norme{\subss{u}{1}}{\infty}\leq 0.5$   &  $\norme{\subss{u}{2}}{\infty}\leq 0.65$   &  $\norme{\subss{u}{3}}{\infty}\leq 0.65$ &   $\norme{\subss{u} {4}}{\infty}\leq 0.55$   &  $\norme{\subss{u}{5}}{\infty}\leq 0.5$    \\
              \hline 
              \multicolumn{6}{c}{}\\
            \end{tabular}
            
            $P_{12} = 4\qquad P_{23}=2\qquad P_{34}=2\qquad P_{45}=3\qquad P_{25}=3$\\                
            
            \caption{Model parameters and constraints for systems $\subss\Sigma i,~i\in1,\ldots,5$.}
            \label{tab:scenario123}
          \end{table}

          \subsection{Scenario 1}
               \label{sec:scenario1}
\begin{figure}[!ht]
                 \centering
                 \includegraphics[scale=0.75]{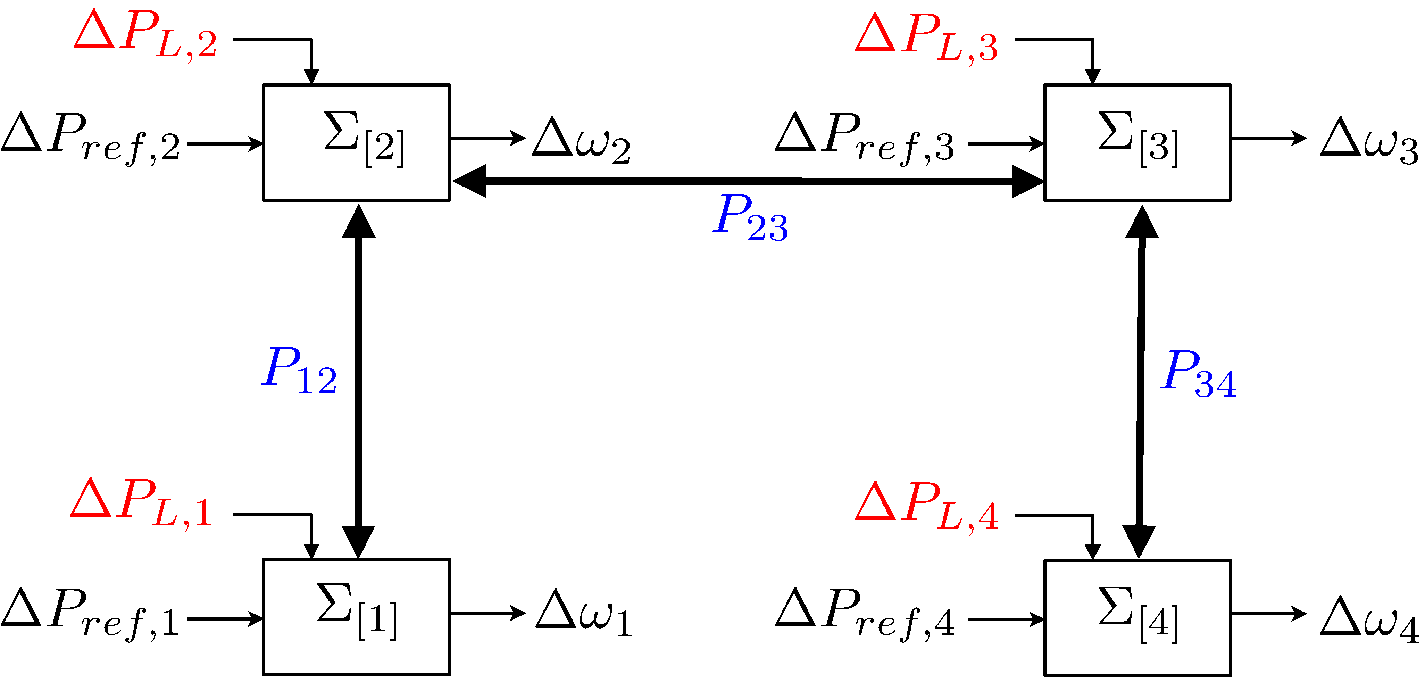}
                 \caption{Power network system of Scenario 1}
                 \label{fig:scenario1}
               \end{figure}

               We consider four areas interconnected as in Figure~\ref{fig:scenario1}.
               We will simulate Scenario 1 using the load steps specified in Table \ref{tab:simulationscen1}.
               \begin{table}[!ht]
                 \centering
                 \begin{tabular}{|c|c|c|}
                   \hline
                   Step time  &  Area $i$ & $\Delta P_{L_i}$ \\
                   \hline
                   5               &      1        &   +0.15             \\
                   \hline
                   15             &      2        &   -0.15             \\
                   \hline
                   20             &      3        &   +0.12             \\
                   \hline
                   40             &      3        &   -0.12             \\
                   \hline
                   40             &      4        &   +0.28            \\
                   \hline
                 \end{tabular}
                 \caption{Load of power $\Delta P_{L_i}$ (p.u.) for simulation in Scenario 1. $+\Delta P_{L_i}$ means a step of required power, hence a decrease of the frequency deviation $\Delta\omega_i$ and therefore an increase of the power reference $\Delta P_{ref_i}$.}
                 \label{tab:simulationscen1}
               \end{table}

          \subsection{Scenario 2}
               \label{sec:scenario2}
\begin{figure}[!ht]
                 \centering
                 \includegraphics[scale=0.75]{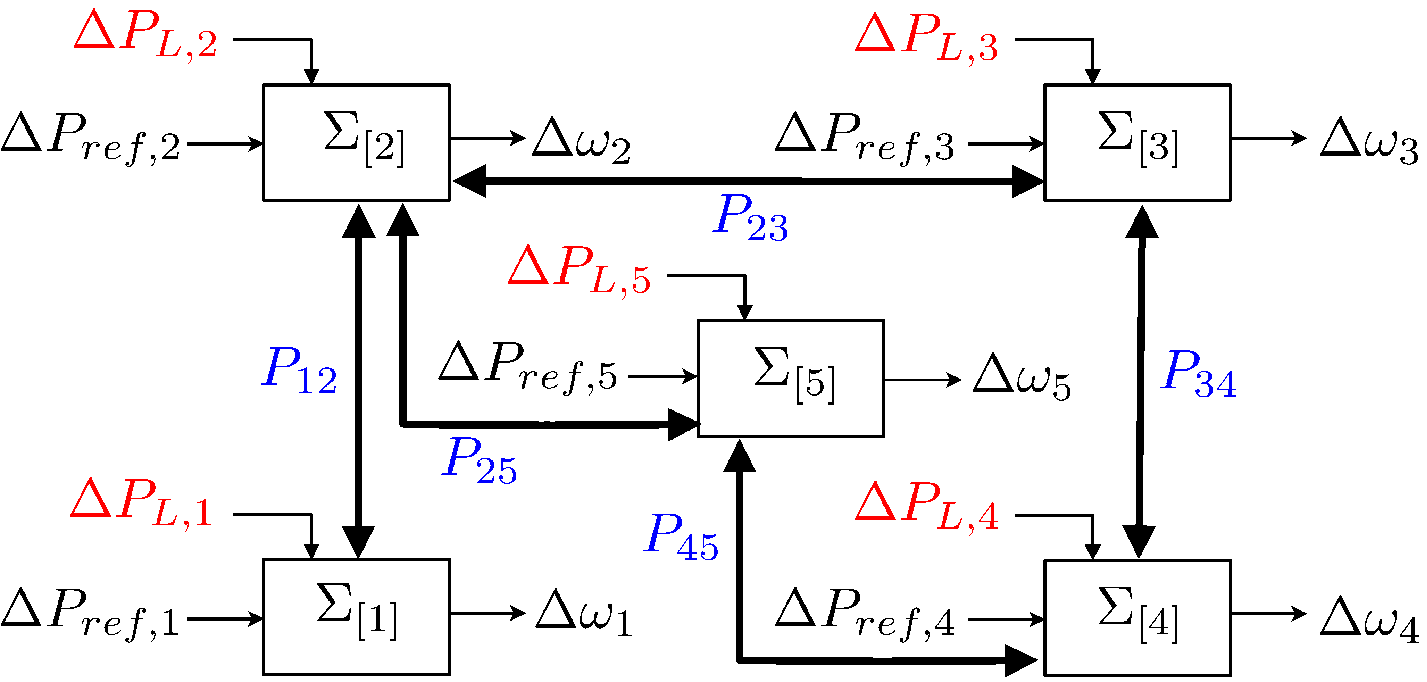}
                 \caption{Power network system of Scenario 2}
                 \label{fig:scenario2}
               \end{figure}

               We consider the power network proposed in Scenario 1 and add a fifth area connected as in Figure \ref{fig:scenario2}.
               We will simulate Scenario 2 using the load steps specified in Table \ref{tab:simulationscen2}.
               \begin{table}[!ht]
                 \centering
                 \begin{tabular}{|c|c|c|}
                   \hline
                   Step time  &  Area $i$ & $\Delta P_{L_i}$ \\
                   \hline
                   5               &      1        &   +0.10             \\
                   \hline
                   15             &      2        &   -0.16             \\
                   \hline
                   20             &      1        &   -0.22             \\
                   \hline
                   20             &      2        &   +0.12             \\
                   \hline
                   20             &      3        &   -0.10             \\
                   \hline
                   30             &      3        &   +0.10             \\
                   \hline
                   40             &      4        &   +0.08             \\
                   \hline
                   40             &      5        &   -0.10             \\
                   \hline
                 \end{tabular}
                 \caption{Load of power $\Delta P_{L_i}$ (p.u.) for simulation in Scenario 2. $+\Delta P_{L_i}$ means a step of required power, hence a decrease of the frequency deviation $\Delta\omega_i$ and therefore an increase of the power reference $\Delta P_{ref_i}$.}
                 \label{tab:simulationscen2}
               \end{table}

          \subsection{Scenario 3}
               \label{sec:scenario3}
\begin{figure}[!ht]
                 \centering
                 \includegraphics[scale=0.75]{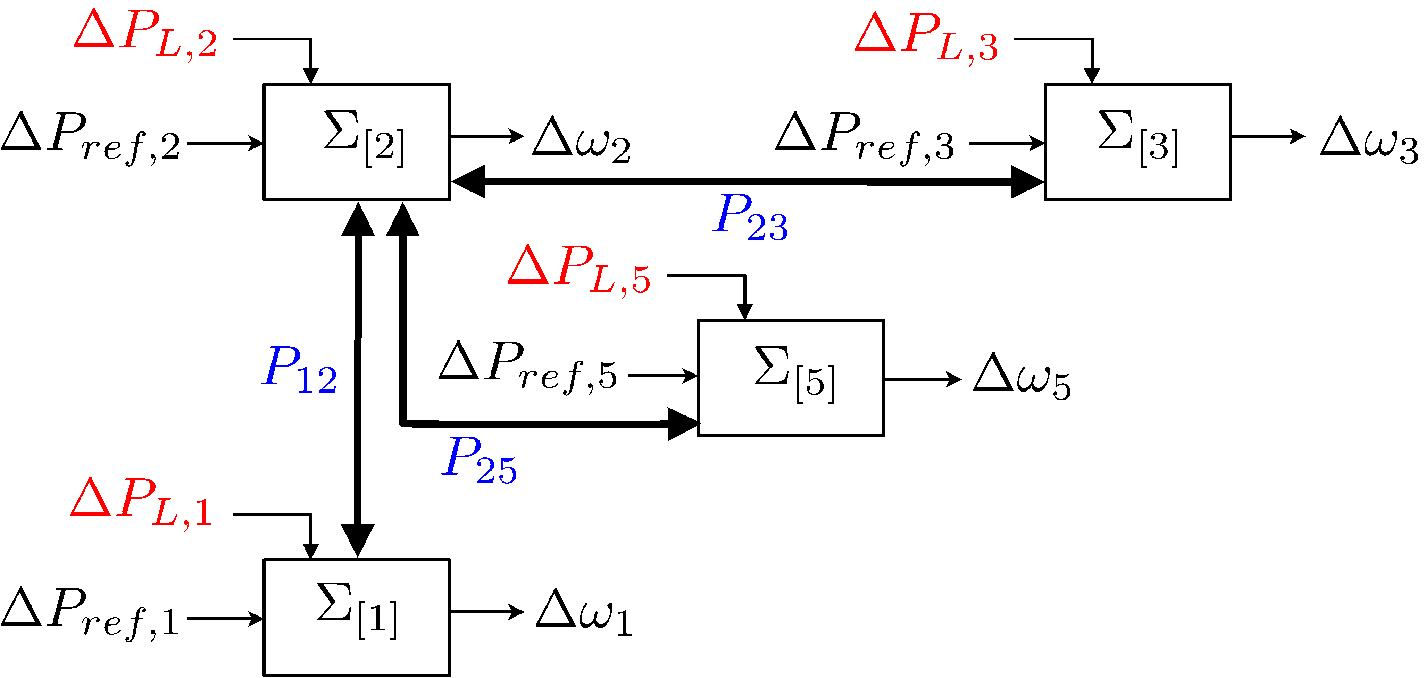}
                 \caption{Power network system of Scenario 3}
                 \label{fig:scenario3}
               \end{figure}

               We consider the power network described in Scenario 2 and disconnect the area $4$, hence obtaining the areas connected as in Figure \ref{fig:scenario3}.               
               We will simulate Scenario 3 using load steps specified in Table \ref{tab:simulationscen3}.
               \begin{table}[!ht]
                 \centering
                 \begin{tabular}{|c|c|c|}
                   \hline
                   Step time  &  Area $i$ & $\Delta P_{L_i}$ \\
                   \hline
                   5               &      1        &   +0.12             \\
                   \hline
                   15             &      2        &   -0.15             \\
                   \hline
                   20             &      5        &   +0.20             \\
                   \hline
                   40             &      2        &   +0.15             \\
                   \hline
                   40             &      3        &   +0.13            \\
                   \hline
                   40             &      5        &   -0.20            \\
                   \hline
                 \end{tabular}
                 \caption{Load of power $\Delta P_{L_i}$ (p.u.) for simulation in Scenario 3. $+\Delta P_{L_i}$ means a step of required power, hence a decrease of the frequency deviation $\Delta\omega_i$ and therefore an increase of the power reference $\Delta P_{ref_i}$.}
                 \label{tab:simulationscen3}
               \end{table}

     \section{Design of the AGC layer for a power network using MPC}
          The goal of the Benchmark is to design the AGC layer for the scenarios introduced in Section \ref{sec:PNS}. Different control schemes will be compared with the centralized MPC scheme described next. For a given Scenario, at time $t$ we solve the centralized optimization problem
          \begin{subequations}   
            \label{eq:MPCProblem}
            \begin{align}              
              &\Pset^N(x (t)): \\
              &\label{eq:costMPCProblem}\min_{\substack{u(t:t+N-1)}}~\sum_{k=t}^{t+N-1}(\norme{x(k)-x^O}{Q}+\norme{u(k)-u^O}{R})+\norme{x(t+N)-x^O}{S})                                        &                    \\
              &\label{eq:dynproblem}x (k+1)=Ax(k)+Bu(k)+L\Delta P_L(t)                                                                                                                            & k\in0:N-1  \\              
              &\label{eq:inhXproblem}x(k)\in\Xset                                                                                                                                                               & k\in0:N-1  \\
              &\label{eq:inVproblem}u(k)\in\Uset                                                                                                                                                                 & k\in0:N-1  \\
              &\label{eq:inTerminalSet}x(N)\in\Xset_f                                                                                                                                                           &
            \end{align}
          \end{subequations}
          and then apply $\Delta P_{ref} = u(0)$. We note that the cost function depend upon $x^O$ and $u^O$ that are defined as $\subss x i^O=(0,~0,~\Delta P_{L_i},~\Delta P_{L_i})$ and $\subss u i^O=\Delta P_{L_i}$. The constraints $\Xset$ and $\Uset$ in \eqref{eq:inhXproblem} and \eqref{eq:inVproblem} are obtained from constraints listed in Table \ref{tab:scenario123}. In the cost function \eqref{eq:costMPCProblem} we set $N=15$, $Q=\diag(Q_1,\ldots,Q_M)$ and $R=\diag(R_1,\ldots,R_M)$, where
          $$
          Q_i = \matr{ 500 & 0 & 0 & 0 \\ 0 & 0.01 & 0 & 0 \\ 0 & 0 & 0.01 & 0 \\ 0 & 0 & 0 & 10 }\mbox{    and    } R_i = 10.
          $$
          Weights $Q_i$ and $R_i$ have been chosen in order to penalize the angular displacement $\Delta\theta_i$ and to penalize slow reactions to power load steps. Since the power transfer between areas $i$ and $j$ is given by
          \begin{equation}
            \label{eq:powerexchanged}
            \Delta P_{{tie}_{ij}}(k) = P_{ij}(\Delta\theta_i(k)-\Delta\theta_j(k))
          \end{equation}
          the first requirement also penalizes huge power transfers.

          In order to guarantee the stability of the closed loop system, we design the matrix $S$ and the terminal constraint set $\Xset_f$ in three different ways.
          \begin{itemize}
          \item\emph{S is full ($MPCfull$)}: we compute the symmetric positive-definite matrix $S$ and the static state-feedback auxiliary control law $K_{aux}x$, by maximizing the volume of the ellipsoid described by $S$ inside the state constraints while fulfilling the matrix inequality $(A+BK_{aux})'S(A+BK_{aux})-S\leq-Q-K_{aux}'RK_{aux}$.
          \item\emph{S is block diagonal ($MPCdiag$)}: we compute the decentralized symmetric positive-definite matrix $S$ and the decentralized static state-feedback auxiliary control law $K_{aux}x$, $K_{aux}=\diag(K_{1},\ldots,K_{M})$ by maximizing the volume of the ellipsoid described by $S$ inside the state constraints while fulfilling the matrix inequality $(A+BK_{aux})'S(A+BK_{aux})-S\leq-Q-K_{aux}'RK_{aux}$.
          \item\emph{Zero terminal constraint ($MPCzero$)}: we set $S=0$ and $\Xset_f = x^O$. 
          \end{itemize}
          
          \subsection{Performance criteria}
               We propose the following performance criteria for evaluating different control schemes.
               \begin{itemize}
               \item\emph{$\eta$-index}
                 \begin{equation}
                   \label{eq:performanceeta}
                   \eta = \frac{1}{T_{sim}}\sum_{k=0}^{T_{sim}-1}\sum_{i=1}^M (\norme{\subss x i(k)-\subss x i^O(k)}{Q_i}+\norme{\subss u i(k)-\subss u i^O(k)}{R_i})
                 \end{equation}
                 where $T_{sim}$ is the time of the simulation. From \eqref{eq:performanceeta}, $\eta$ is a weighted average of the error between the real state and the equilibrium state and between the real input and the equilibrium input.
               \item \emph{$\Phi$-index}
                 \begin{equation}
                   \label{eq:performancePhi}
                   \Phi = \frac{1}{T_{sim}}\sum_{k=0}^{T_{sim}-1}\sum_{i=1}^M\sum_{j\in\NN_i}\abs{\Delta P_{{tie}_{ij}}(k)}T_s
                 \end{equation}
                 where $T_{sim}$ is the time of the simulation and $\Delta P_{{tie}_{ij}}$ is the power transfer between areas $i$ and $j$ defined in \eqref{eq:powerexchanged}. This index gives the average power transferred between areas. In particular, if the $\eta$-index is equal for two regulators, the best controller is the one that has the lower value of $\Phi$.
               \end{itemize}

     \section{Control Experiments}
          We applied the centralized MPC schemes introduced in the previous section to scenarios 1, 2 and 3. Furthermore, for each scenario we discretized the continuos system with both discretization schemes $D$ and $Dss$. At time $t$ we solve the optimization problem \eqref{eq:MPCProblem} and then apply the control action to the continuos-time system, keeping the value constant between time $t$ and $t+1$. If at time $t$ the power load increases or decreases, we assume the controller can use this information at time $t$. This means at time $t$ the controller knows exactly the value of $\Delta P_L$ hence can use it. We highlight that violation of this assumption can impact considerably on the index $\eta$. In all experiments we use $T_{sim}=100$. In Table \ref{tab:simulationsEta} and \ref{tab:simulationsPhi} the values of the performance parameters $\eta$ and $\Phi$, respectively, are reported for each control experiment. 

          \begin{table}[!ht]
            \centering
            \begin{tabular}{|c||c|c||c|c||c|c|}
              \hline
                                     & \multicolumn{2}{|c||}{Scenario 1} & \multicolumn{2}{|c||}{Scenario 2} & \multicolumn{2}{|c|}{Scenario 3} \\
              \hline
                                     &          $D$            &    $Dss$       &          $D$           &    $Dss$       &          $D$          &    $Dss$        \\
              \hline
              $MPCfull$      &       0.0249           &   0.0249        &    0.0346            &   0.0347        &    0.0510           &      0.0511       \\
              \hline
              $MPCdiag$     &      0.0249            &    0.0249      &     0.0346            &   0.0347        &    0.0510           &    0.0511         \\
              \hline
              $MPCzero$     &       0.0249           &   0.0249       &    0.0346             &   0.0347        &   0.0510            &   0.0511            \\
              \hline
            \end{tabular}
            \caption{Values of the performance parameter $\eta$ using different centralized MPC schemes for the AGC layer.}
            \label{tab:simulationsEta}
          \end{table}

          \begin{table}[!ht]
            \centering
            \begin{tabular}{|c||c|c||c|c||c|c|}
              \hline
                                     & \multicolumn{2}{|c||}{Scenario 1} & \multicolumn{2}{|c||}{Scenario 2} & \multicolumn{2}{|c|}{Scenario 3} \\
              \hline
                                     &          $D$            &    $Dss$       &          $D$           &    $Dss$       &          $D$          &    $Dss$        \\
              \hline
              $MPCfull$       &        0.0030         &   0.0029       &    0.0063              &     0.0060     &      0.0060          &   0.0058         \\
              \hline
              $MPCdiag$     &       0.0030          &   0.0029        &   0.0063              &   0.0061        &   0.0060             &    0.0058        \\
              \hline
              $MPCzero$     &    0.0030            &     0.0028       &    0.0063              &  0.0059        &  0.0059              &   0.0058          \\
              \hline
            \end{tabular}
            \caption{Values of the performance parameter $\Phi$ using different centralized MPC schemes for the AGC layer.}
            \label{tab:simulationsPhi}
          \end{table}

     \section{Supporting Matlab files}

         We provide the Matlab files for the parameters in Table \ref{tab:scenario123} (\mbox{parameters.m}) and for all control experiments. Each file $.mat$ of the control experiments contains 
         \begin{itemize}
         \item the matrices of the continuos linear system ($Ac$, $Bc$, $Cc$, $Dc$, $Lc$);
         \item the matrices of the discretized linear system ($A$, $B$, $C$, $D$, $L$, $Ts$);
         \item parameters of the controller ($Q$, $R$, $S$, $N$, $xO$, $uO$);
         \item parameters of the control experiment  $Tsim$ and $deltaPload$, where $deltaPload$ corresponds to $\Delta P_L$;
         \item the results of the control experiment $x$, $deltaPref$, $\eta$ and $\Phi$, where $deltaPref$ corresponds to $\Delta P_{ref}$.
         \end{itemize}
         
         For each Scenario we included also a Simulink model. In particular, one can load the file $.mat$ of a control experiment and simulate the power network system given the power load steps and the power reference computed through centralized MPC.

          \subsection{Example of simulation}
               In the following we illustrate how to use the files $.mat$ and the Simulink models through an example. Assume we want to simulate Scenario 2 using the discretization $Dss$ and centralized MPC with zero terminal constraint ($MPCzero$). In the folder of each scenario there are six folders labeled as \emph{[discretization scheme]\_[mpc type]}. Hence, we have to use files in folder \emph{Dss\_MPCzero}. In this folder we can find the data of the required control experiment as \emph{dataSim.mat}. The previous operations are performed with the Matlab commands:
               \begin{lstlisting}
                 cd scenario2
                 load Dss_MPCzero/dataSim
               \end{lstlisting}
               We can simulate different scenarios using the Simulink models present in the folder of each scenario. For Scenario 2 we then open the file \emph{simulatorPNS\_AGC\_2.mdl}. Start a simulation from Simulink will produce the results of the control experiments. These steps are performed with the Matlab commands:
               \begin{lstlisting}
                 open('simulatorPNS_AGC_2')
                 sim('simulatorPNS_AGC_2')
               \end{lstlisting}

     \section{Benchmark exercise}
          The aim is to design decentralized/distributed controllers for the scenarios described in Section \ref{sec:PNS}.

          Depending on the control technique adopted either $D$ or $Dss$ discretization schemes can be chosen. 

          The first goal of a distributed AGC layer is to have performance in terms of $\eta$ similar to centralized MPC. Matching also the values of $\Phi$ can be seen as a secondary objective.

          Alternative control schemes will be also ranked according to the degree of decentralization of the design process. Ideally, the controller of each area should be designed independently of the others and using information from a limited number of other areas. Decentralized design is important in PNS because if an area needs to be isolated or a new area is plugged into the network one would like to avoid the redesign the whole AGC layer and rather retune just a limited number of local controllers in order to guarantee asymptotic stability and constraints satisfaction for the whole network.

     \bibliographystyle{alpha}
     \bibliography{Hycon2_Benchmark_PNS}

\end{document}